\documentclass[conference]{IEEEtran}
\IEEEoverridecommandlockouts

\usepackage{amsmath}
\usepackage{amsfonts}
\usepackage{amssymb}
\usepackage{graphicx}
\usepackage{epsfig}
\usepackage[scriptsize,center]{caption}
\usepackage{subfig}
\usepackage{float}
\usepackage{epstopdf}
\usepackage{array}
\usepackage{multirow}
\usepackage{setspace}
\usepackage{color}
\usepackage[section]{placeins}
\usepackage{url}
\usepackage{algorithm}
\usepackage{algpseudocode}
\usepackage{multicol}
\usepackage{soul}
\usepackage{comment}
\usepackage[acronym]{glossaries}
\usepackage[normalem]{ulem}

\title{NR-U and WiGig Coexistence in 60 GHz Bands}
\author{
  \IEEEauthorblockN{
    Natale~Patriciello\IEEEauthorrefmark{1},~Sanjay~Goyal\IEEEauthorrefmark{2},~Sandra~Lagen\IEEEauthorrefmark{1},~Lorenza~Giupponi\IEEEauthorrefmark{1},~Biljana~Bojovi\'c\IEEEauthorrefmark{1},~Alpaslan~Demir\IEEEauthorrefmark{2},\\ ~Mihaela~Beluri\IEEEauthorrefmark{2}
   }
    \IEEEauthorblockA{\IEEEauthorrefmark{1}Centre Tecnol\`ogic de Telecomunicacions de Catalunya (CTTC/CERCA), Castelldefels, Barcelona, Spain\\}
	\IEEEauthorblockA{\IEEEauthorrefmark{2}InterDigital Communications, Inc., Melville, New York, USA \\
	\{natale.patriciello, sandra.lagen, lorenza.giupponi, biljana.bojovic\}@cttc.es, \\ \{sanjay.goyal, alpaslan.demir, mihaela.beluri\}@interdigital.com}
}

\begin{document}
\maketitle

\begin{abstract}
In December 2019, the 3GPP defined the road-map for Release-17, which includes new features on the operation of New Radio (NR) in millimeter-wave bands with highly directional communications systems, i.e., up to 52.6 GHz.
In this paper, a system-level simulation based study on the coexistence of NR-based access to unlicensed spectrum (NR-U) and an IEEE technology, i.e., 802.11ad Wireless Gigabit (WiGig), at 60 GHz bands is conducted. For NR-U, an extension of NR Release-15 based model is used such that the 60 GHz regulatory requirements are satisfied.
First, the design and capabilities of the developed open source ns-3 based simulator are presented and then end-to-end performance results of coexistence with different channel access mechanisms for NR-U in a 3GPP indoor scenario are discussed. It is shown that NR-U with Listen-Before-Talk channel access mechanism does not have any adverse impact on WiGig performance in terms of throughput and latency, which demonstrates that NR-U design fulfills the fairness coexistence objective, i.e., NR-U and WiGig coexistence is proven to be feasible.
\end{abstract}

\IEEEpeerreviewmaketitle

\newacronym{scs}{SCS}{SubCarrier Spacing}
\newacronym{ofdm}{OFDM}{Orthogonal Frequency Division Multiplexing}
\newacronym{fdm}{FDM}{Frequency Division Multiplexing}
\newacronym{cpu}{CPU}{Central Processing Unit}
\newacronym{tcp}{TCP}{Transmission Control Protocol}
\newacronym{tcpw}{TCPW}{TCP Wave}
\newacronym{lte}{LTE}{Long Term Evolution}
\newacronym{nr}{NR}{New Radio}
\newacronym{nru}{NR-U}{NR-based access to Unlicensed spectrum}
%\newacronym{cca}{CCA}{Congestion Control Algorithm}
\newacronym{cwnd}{cWnd}{Congestion Window}
\newacronym{caa}{CAA}{Congestion Avoidance Algorithm}
\newacronym{bbr}{BBR}{Bottleneck Bandwidth and Round-trip propagation time}
\newacronym{nv}{NV}{New Vegas}
\newacronym{rtt}{RTT}{Round-Trip Time}
\newacronym{ietf}{IETF}{Internet Engineering Task Force}
\newacronym{rfc}{RFC}{Request For Comments}
\newacronym{gcc}{GCC}{GNU Compiler Collection}
\newacronym{tso}{TSO}{TCP Segmentation Offloading}
\newacronym{tsq}{TSQ}{TCP Small Queue}
\newacronym{gbr}{GBR}{Guaranteed Bit Rate}
\newacronym{nongbr}{non-GBR}{non-Guaranteed Bit Rate}
\newacronym{enb}{eNB}{Evolved Node B}
\newacronym{dpi}{DPI}{Deep Packet Inspection}
\newacronym{rlc}{RLC}{Radio Link Control}
\newacronym{bsr}{BSR}{Buffer Status Report}
\newacronym{qos}{QoS}{Quality of Service}
\newacronym{aqm}{AQM}{Active Queue Management}
\newacronym{rds}{RDS}{Radio Data Scheduler}
\newacronym{tc}{TC}{Traffic Control}
\newacronym{drb}{DRB}{Data Radio Bearer}
\newacronym{rnti}{RNTI}{Radio Network Temporary Identifier}
\newacronym{bql}{BQL}{Byte Queue Limits}
\newacronym{ue}{UE}{User Equipment}
\newacronym{am}{AM}{Acknowledged Mode}
\newacronym{epc}{EPC}{Evolved Packet Core}
\newacronym{cn}{CN}{Core Network}
\newacronym{gnb}{gNB}{next-Generation Node B}
\newacronym{ran}{RAN}{Radio Access Network}
\newacronym{3gpp}{3GPP}{3rd Generation Partnership Project}
\newacronym{5g}{5G}{fifth Generation}
\newacronym{dl}{DL}{DownLink}
\newacronym{ul}{UL}{UpLink}
\newacronym{tti}{TTI}{Transmission Time Interval}
\newacronym{sr}{SR}{Scheduling Request}
\newacronym{bsr}{BSR}{Buffer Status Report}
\newacronym{e2e}{E2E}{End-To-End}
\newacronym{embb}{eMBB}{enhanced Mobile BroadBand}
\newacronym{urllc}{URLLC}{Ultra-Reliable and Low-Latency Communications}
\newacronym{mmtc}{mMTC}{massive Machine Type Communications}
\newacronym{ul}{UL}{UpLink}
\newacronym{dl}{DL}{DownLink}
\newacronym{phy}{PHY}{Physical layer}
\newacronym{mac}{MAC}{Medium Access Control}
\newacronym{prb}{PRB}{Physical Resource Block}
\newacronym{pps}{pps}{Packets Per Second}
\newacronym{cp}{CP}{Cyclic Prefix}
\newacronym{tbs}{TBS}{Transport Block Size}
\newacronym{tb}{TB}{Transport Block}
\newacronym{cb}{CB}{Code Block}
\newacronym{gtp}{GTP}{GPRS Tunneling Protocol}
\newacronym{sap}{SAP}{Service Access Point}
\newacronym{tm}{TM}{Transparent Mode}
\newacronym{um}{UM}{Unacknowledged Mode}
\newacronym{am}{AM}{Acknowledged Mode}
\newacronym{sm}{SM}{Saturation Mode}
\newacronym{sinr}{SINR}{Signal-to-Interference-plus-Noise Ratio}
\newacronym{rrc}{RRC}{Radio Resource Control}
\newacronym{tdma}{TDMA}{Time-Division Multiple Access}
\newacronym{ofdma}{OFDMA}{Orthogonal Frequency-Division Multiple Access}
\newacronym{rbg}{RBG}{Resource Block Group}
\newacronym{rb}{RB}{Resource Block}
\newacronym{dci}{DCI}{Downlink Control Information}
\newacronym{uci}{UCI}{Uplink Control Information}
\newacronym{ipat}{IPAT}{Inter-Packet Arrival Time}
\newacronym{pdsch}{PDSCH}{Physical Downlink Shared Channel}
\newacronym{pusch}{PUSCH}{Physical Uplink Shared Channel}
\newacronym{pucch}{PUCCH}{Physical Uplink Control Channel}
\newacronym{pdcch}{PDCCH}{Physical Downlink Control Channel}
\newacronym{tdd}{TDD}{Time Division Duplex}
\newacronym{fdd}{FDD}{Frequency Division Duplex}
\newacronym{rach}{RACH}{Random Access Channel}
\newacronym{cbr}{CBR}{Constant Bit Rate}
\newacronym{los}{LoS}{Line-of-Sight}
\newacronym{mcs}{MCS}{Modulation Coding Scheme}
\newacronym{bwp}{BWP}{Bandwidth Part}
\newacronym{cqi}{CQI}{Channel Quality Indicator}
\newacronym{bler}{BLER}{Block Error Rate}
\newacronym{tbler}{TBLER}{Transport Block Error Rate}
\newacronym{mi}{MI}{Mutual Information}
\newacronym{l2sm}{L2SM}{Link to System Mapping}
\newacronym{sliv}{SLIV}{Start and Length Indicator Value}
\newacronym{mmwave}{mmWave}{millimeter-wave}
\newacronym{pdu}{PDU}{Packet Data Unit}
\newacronym{ca}{CA}{Carrier Aggregation}
\newacronym{snr}{SNR}{Signal-to-Noise Ratio}
\newacronym{sinr}{SINR}{Signal to Interference-plus-Noise Ratio}
\newacronym{pdcp}{PDCP}{Packet Data Convergence Protocol}
\newacronym{sdap}{SDAP}{Service Data Adaptation Protocol}
\newacronym{sdu}{SDU}{Service Data Unit}
\newacronym{nas}{NAS}{Non-Access Stratum}
\newacronym{sme}{SME}{Small and Medium Enterprise}
\newacronym{rat}{RAT}{Radio Access Technology}
\newacronym{pgw}{PGW}{Packet data network GateWay}
\newacronym{sgw}{SGW}{Service GateWay}
\newacronym{ldpc}{LDPC}{low Density Parity Check}
\newacronym{cca}{CCA}{Clear Channel Assessment}
\newacronym{csmaca}{CSMA/CA}{Carrier Sense Multiple Access with Collision Avoidance}
\newacronym{ccm}{CCM}{Component Carrier Manager}
\newacronym{cam}{CAM}{Channel Access Manager}
\newacronym{cws}{CWS}{Contention Window Size}
\newacronym{ed}{ED}{Energy detection}
\newacronym{harq}{HARQ}{Hybrid Automatic Repeat Request}
\newacronym{ap}{AP}{Access Point}
\newacronym{laa}{LAA}{Licensed-Assisted Access}
\newacronym{lbt}{LBT}{Listen-Before-Talk}
\newacronym{mcot}{MCOT}{Maximum Channel Occupancy Time}
\newacronym{cot}{COT}{Channel Occupancy Time}
\newacronym{rar}{RAR}{Random Access Response}
\newacronym{ip}{IP}{Internet Protocol}
\newacronym{ss/pbch}{SS/PBCH}{Synchronization Signal/Physical Broadcast Channel}
\newacronym{prach}{PRACH}{Physical Random Access Channel}

\section{Introduction}
\label{sec:intro}
The \gls{3gpp} has recently completed the first phase of the standardization of a new Radio
Access Technology (RAT) for the 5th Generation (5G) systems, i.e., 3GPP \gls{nr}~\cite{TS38300}.
One of the main new features of \gls{nr} is the built-in support for carrier frequencies up to 52.6 GHz. The \gls{mmwave} frequencies have a much larger spectrum availability than the congested sub-6 GHz bands~\cite{pi:11}, which offers {much} higher data rates than in previous generations of mobile communication systems.

{Additionally}, \gls{nr} is being designed with a native feature to operate in unlicensed spectrum through the so-called \gls{nru} extension~\cite{lagen:19}. Differently from LTE Licensed-Assisted Access (LTE-LAA)~\cite{kwon:17} and LTE-Unlicensed (LTE-U)~\cite{LTEUQCom}, which only work on a carrier aggregated to the licensed band, \gls{nru} design considers dual connectivity and standalone operation in unlicensed bands as well, which is an unprecedented milestone for cellular systems to achieve. The design of \gls{nru} started in a Release-16 Study Item in 2018, which led to TR 38.889~\cite{TR38889}, and is currently under development in a  Release-16 Work Item. The target of current Work Item is the unlicensed/shared spectrum for sub-7 GHz bands (including the 2.4, 3.5, 5, and 6 GHz bands). However, for Release-17, a study item has been just approved to extend NR operation up to 71 GHz, and so also in unlicensed 60 GHz~\cite{RP-193259}\cite{RP-193229}. In 60 GHz bands, {9 GHz and 14 GHz spectrum have recently been released in Europe and in the USA, respectively}, which provide 10$\times$ and 16$\times$ times, respectively, {more} unlicensed spectrum {compared to} sub-7 GHz bands. 

In \gls{mmwave} unlicensed bands, \gls{nru}  needs to coexist with IEEE 802.11ad (also known as WiGig)~\cite{6979964} and its successor IEEE 802.11ay~\cite{8088544}. Hence, the research in this domain to design an efficient multi-\gls{rat} coexistence at unlicensed bands {is expected to} grow exponentially, which requires the availability of complex high fidelity simulation tools for evaluation of coexistence performance. The results presented in TR 38.889~\cite{TR38889} on \gls{nru} and Wi-Fi coexistence in the 5 GHz band are obtained by multiple companies, through simulators that are not usually publicly available. Such results are not easily reproducible, and system performance metrics included therein are presented without much detail revealed about the underlying {models/assumptions}. For this purpose, we focus on extending {an} open source ns-3 simulator, which offers opportunities for reproducible research and collaborative development, with high fidelity, full stack and end-to-end models of both 3GPP (LTE~\cite{Bojovic2019} and NR~\cite{PATRICIELLO2019101933}) and IEEE technologies (Wi-Fi and WiGig~\cite{hany:16}). In particular, in this paper, we focus on the NR-U and WiGig coexistence in 60 GHz bands.

The first contribution of this paper is the presentation of an open-source extension of the ns-3 simulator\footnote{The software is available from https://5g-lena.cttc.es/} targeting (\textit{i}) \gls{nru}  models, and (\textit{ii}) \gls{nru} and WiGig coexistence  in  \gls{mmwave}   bands, 
using the recently released models for NR~\cite{PATRICIELLO2019101933} and WiGig~\cite{hany:16}. To develop the NR-U model, we extended an NR Release-15 based model~\cite{PATRICIELLO2019101933} by incorporating the 60 GHz regulatory requirements. Our objective is to study coexistence scenarios between 3GPP and IEEE technologies in the unlicensed \gls{mmwave} spectrum from an end-to-end perspective (including e.g., multiple users, multiple application flows, complex deployment scenarios).
The second contribution of the paper is a high-level study of channel occupancy, latency, and throughput of an indoor 3GPP-oriented scenario, where a standalone deployment of \gls{nru}  coexists with WiGig. We {focus} on the \gls{lbt} protocol that is currently being considered by {the} 3GPP~\cite{TR38889} as the most significant feature to allow a fair coexistence between different technologies. {We also include} other {channel access}  schemes for NR-U that may gain momentum in the future, {i.e.,} (\textit{i}) a duty-cycle version (inspired by LTE-U), which does not perform a \gls{cca} before transmission and {may be suitable} in world regions without LBT requirements (like in the USA), and (\textit{ii}) an always-on variant (i.e., uninterrupted NR) that is evaluated as a benchmark. Our results show that \gls{nru}  and WiGig at 60 GHz band fairly coexist with each other, when \gls{nru}  incorporates a LBT or duty-cycle {based} access. 

The rest of the paper is organized as follows. Section~\ref{sec:nr3gpp} reviews {3GPP NR-U activities.}  Section~\ref{sec:simulator} presents our \gls{nru} model.  Section~\ref{sec:uses} discusses {the results of the end-to-end performance evaluation.}  Section~\ref{sec:conc} concludes the paper.
Throughout this paper, in line with \gls{3gpp} and IEEE terminologies, we refer to an \gls{nru} base station, an \gls{nru} terminal, a WiGig base station, and a WiGig terminal as \gls{gnb}, \gls{ue}, \gls{ap}, and station (STA), respectively. 

\section{NR-U in 3GPP}
\label{sec:nr3gpp}
In December 2018, {the} 3GPP approved a new Release-16 Work Item to include support for NR-U in the unlicensed sub-7 GHz bands,
as a follow up of the previous Release-16 Study Item that resulted in TR 38.889~\cite{TR38889}.
{The} 3GPP has defined three main deployment scenarios for NR-U:
\begin{itemize}
\item  Carrier Aggregation, which is based on the previous design of LTE-LAA in Release-13~\cite{kwon:17}. 
\item  Dual Connectivity, which is based on the previous design of LTE-eLAA in Release-14. 
\item Standalone, which is a novel approach in Release-16. 
\end{itemize}
In the first two modes, NR-U can be anchored to both licensed LTE and NR. In standalone NR-U, similar to an approach taken by MulteFire Alliance {for} standalone operation of LTE in unlicensed bands~\cite{8493138}, NR-U is expected to work in unlicensed spectrum without being anchored to any licensed carrier. 

The objective of {the} 3GPP is to define the necessary enhancements to NR to determine a single global solution for NR-U. The key basis for all the enhancements is to be compliant with the regulatory requirements~\cite{lagen:19}, such as \gls{lbt}, maximum \gls{cot}, occupied channel bandwidth (OCB), and power limits. 
The LBT procedure{~\cite{Bojovic2019}} is defined as a mechanism for a \gls{cca}  check  before  using  the  channel and is a regulatory requirement for unlicensed bands in  Europe  and  Japan. 
The requirement of LBT creates uncertainty for the channel availability, which is fundamentally different from the licensed-based access, where the transmissions occur at pre-scheduled fixed times. Based on that, modifications to several Release-15 NR features are being considered, including:
\begin{itemize}
\item Initial access, e.g., {changes to} \gls{ss/pbch}
transmissions, random-access procedure, and preamble transmissions due to LBT and OCB requirements
\item Downlink channels and signals, e.g., increase flexibility in {control and data channel} transmissions due to LBT
\item Uplink channels and signals, e.g., interlaced based 
design for uplink channels due to OCB, flexible starting point due to LBT. 
\item Paging, e.g., flexibility in monitoring paging signal due to LBT.
\item \gls{harq} procedures, e.g., additional ACK/NACK transmission opportunities. 
\item Configured grants, e.g., flexibility in time-domain resource allocation. 
\item Wide band operations, e.g., to support transmission of bandwidth larger than {a Wi-Fi channel bandwidth of} 20 MHz.
\item Measurement framework, e.g., changes to radio link monitoring procedure due to LBT. 
\end{itemize}

Additionally, different \gls{lbt}-based channel access procedures, defined by the LBT category and the corresponding parameters, are discussed for each of the downlink and uplink channels under different conditions (e.g., gNB-initiated COT or UE-initiated COT). In particular, four \gls{lbt} categories have been defined for NR-U~\cite{TR38889}, which are inspired by Wi-Fi CSMA/CA 
and which were also standardized for LTE-LAA~\cite{Bojovic2019}:
\begin{itemize}
\item	Category 1 (Cat1 \gls{lbt}): immediate transmission (i.e., no \gls{lbt}).
\item	Category 2 (Cat2 \gls{lbt}): \gls{lbt} without random back-off.
\item	Category 3 (Cat3 \gls{lbt}): \gls{lbt} with random back-off and fixed contention window.
\item	Category 4 (Cat4 \gls{lbt}): \gls{lbt} with random back-off and exponential contention window.
\end{itemize}

As we mentioned earlier, the current 3GPP Release-16 work item considers the design only for sub-7 GHz unlicensed bands. 
The design for above 7 GHz unlicensed bands will be part of the Release-17, based on recent 3GPP agreements to scale up NR to 52.6-71 GHz, including licensed and unlicensed bands~\cite{RP-193259,RP-193229}. 
The key design aspects related to the coexistence of NR-U at 60 GHz unlicensed band, such as directional LBT and corresponding beam management impacts~\cite{lagen:19}, are expected to be considered as part of this item.

In terms of non-3GPP efforts, MulteFire Alliance recently approved a study item on standalone NR-U operation at 60 GHz band in June 2019, where a feasibility study (e.g., waveform, channel models, etc.) is expected to be conducted.
Also, we envision that other initiatives may appear around the world for NR operation in unlicensed spectrum that do not rely on \gls{lbt} but rather consider alternative channel access procedures, particularly for such regions in which \gls{lbt} is not mandatory, as already happened with LTE-U in the USA~\cite{LTEUQCom}.
Indeed, depending on the situations and environments, maybe no additional features are added on top of NR to make it operational in unlicensed mmWave bands, because, in some situations, due to the {directional} transmissions, interference could be negligible.

\section{NR-U Simulation Models for Coexistence}
\label{sec:simulator}
We designed an NR-U and WiGig coexistence simulator by integrating the enhanced versions of the ns-3 NR~\cite{PATRICIELLO2019101933} and the WiGig~\cite{hany:16} models. 
In the WiGig model~\cite{hany:16}, we made changes so that (i) interference from other RATs can be modeled and taken into account, and (ii) channel models, antenna models {for uniform planar arrays, and element radiation patterns} are compliant with {the} 3GPP {recommendations for above 6 GHz~\cite{TR38900}.} 
We fixed these aspects to be able to simulate the 3GPP scenarios. We also unified the beamforming representation through antenna weights (a.k.a. beamforming vectors), instead of spatial radiation patterns which were used in the WiGig model, to be compatible with the 3GPP channel model (based on channel matrices) and to enable interaction of the two RATs. 

\begin{figure}[!t]
  \centering
  \includegraphics[width=1\linewidth]{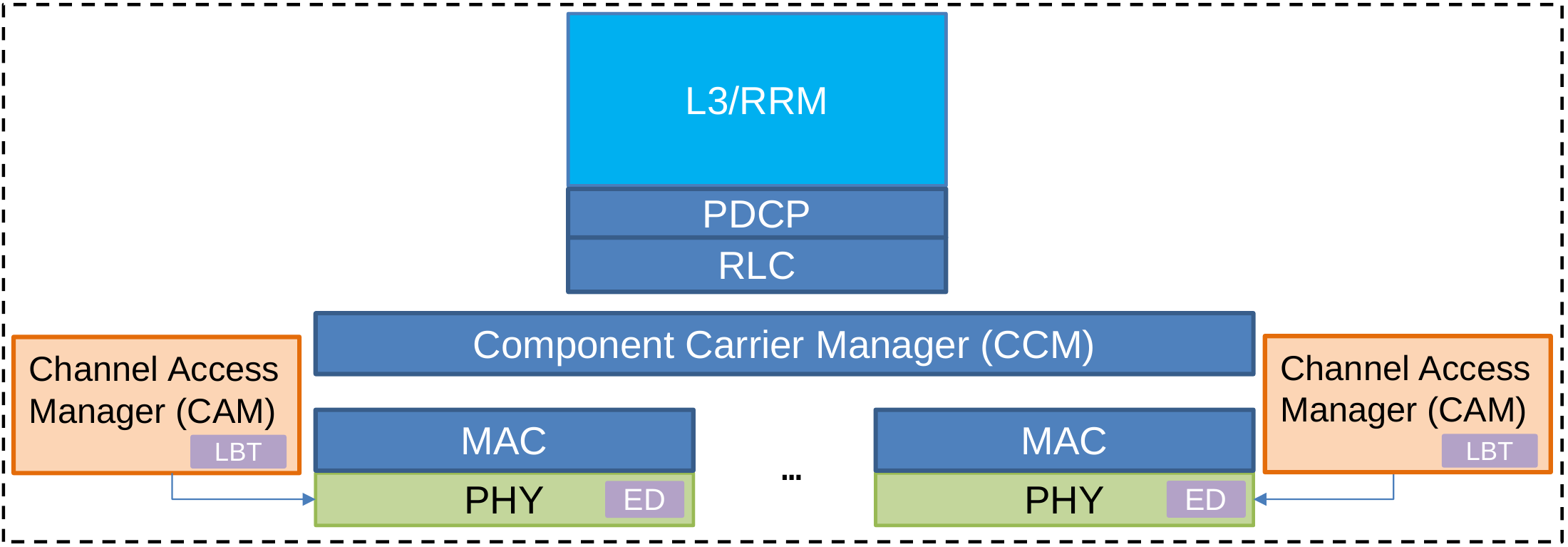}
  \caption{NR-U device architecture with multiple component carriers.}
  \label{fig:architecture}
  \vspace{-0.2cm}
\end{figure}

In the NR model~\cite{PATRICIELLO2019101933}, we incorporated the distinguishing features of NR-U.
As the standardization works for NR/NR-U above 52.6 GHz have not yet started, to build the NR-U model we used an NR Release-15 design~\cite{PATRICIELLO2019101933} and extended it to incorporate the 60 GHz regulatory requirements of maximum COT, \gls{lbt}, OCB, and power limits.
Fig.~\ref{fig:architecture}  presents the architecture of our NR-U device implementation design. A \gls{ccm} manages the traffic distribution among different carriers. 
For each carrier, we have a \gls{cam} that models the presence and the type of the \gls{lbt} algorithm. The sensing capability is incorporated to the NR-U model through the \gls{ed} block at the \gls{phy}, which performs \gls{cca} based on indication from the \gls{lbt} block in the \gls{cam} with the ultimate goal of checking channel availability before transmitting on it. In this work, we implemented omnidirectional sensing, i.e., omni-\gls{lbt}, at the \gls{gnb}s. On the \gls{ue} side, instead, we focused on directional sensing, i.e., dir-\gls{lbt}, since differently from the \gls{gnb}, the \gls{ue} only has to communicate with its \gls{gnb}.

We implemented the 3GPP \gls{lbt} procedure and all four \gls{lbt} categories as subclasses (see description in Section~\ref{sec:nr3gpp}). 
In the downlink, all the \gls{lbt} categories are supported, while in the uplink, it is possible to use only Cat1 and Cat2 \gls{lbt} {because only} downlink data transmission {is simulated}. Every time when an \gls{lbt} is successful, the channel is granted for the maximum \gls{cot} duration. All the LBT categories have different attributes to configure: the \gls{ed} threshold, the CCA slot duration, the deferral interval during CCA, and the maximum COT duration. In addition, the simulator allows configuration of: the minimum {and the maximum values of} \gls{cws} for Cat4 LBT, the \gls{cws} for Cat3 LBT, and the deferral period for Cat2 LBT.
The values that we use for simulations are reported in the next section.

An OnOff \gls{cam} is also implemented, which has a duty-cycle based behavior under which it alternates between ON and OFF periods, without performing LBT to access the channel. 
An UE is assumed to be synchronized with its gNB such that both use the same duty cycle pattern. In the following, we will use the terminology ``AlwaysOn'' to indicate the Cat1 LBT based channel access, in which NR-U operates in an uninterrupted fashion.

Another notable design aspect is the decision when LBT has to be performed with respect to the \gls{mac} processing. There are two options:
\begin{enumerate}
 \item Start the LBT procedure before the \gls{mac} starts the scheduling decisions (hence, passing the data to the \gls{mac} scheduler only after the channel has been declared clear);
 \item Start the LBT procedure after the \gls{mac} has processed and scheduled the data (therefore, assessing the channel already knowing the frame structure that the \gls{phy} must send).
\end{enumerate}
As the \gls{mac} works ahead of the slot in which the data occupies the channel, e.g., 4 slots (2 ms) in LTE, therefore, the two options are not equivalent. Option (1) may generate an inefficiency in spectrum usage because there is a gap between when the channel is granted and when it gets occupied. On the other hand, in option (2), there is a risk that the channel is not granted when the scheduler has decided to occupy it. In our implementation, we opted to reduce the inefficiency in channel occupancy, and therefore selected the option (2). Option (2) also guarantees that the implementation is adequate for sub-7 GHz bands, where the duration of the slot is higher than the one considered for mmWave bands, and consequently the inefficiency due to option (1) would be significant.

Finally, to meet the OCB requirement, we use a \gls{tdma} beam-based access, in which OFDM symbols are allocated among UEs, and spread control channels through the whole bandwidth. Regarding power limits, we meet the maximum radiated power limit and distribute the power uniformly among the physical resource blocks to meet the   spectral power density limit.

The proposed NR-U device model supports Carrier Aggregation NR-U and Standalone NR-U deployment scenarios, as described in Section~\ref{sec:nr3gpp}. 
It does not include the enhancements related to {initial access procedure, downlink/uplink control/data channels} that are being considered in the 3GPP for sub-7 GHz unlicensed bands (as discussed in Section~\ref{sec:nr3gpp}), however, as discussed above, it satisfies the requirements of LBT, maximum COT, and OCB. All in all, the developed NR-U model provides the basis for NR to operate in unlicensed mmWave spectrum while meeting the regulatory requirements, and new features may
easily  be  incorporated  in the future as the specification proceeds.

\section{The simulator in action}
\label{sec:uses}
\subsection{Simulation scenario}
\label{sec:uses_a}
\begin{figure}[!t]
  \centering
  \includegraphics[width=1\linewidth]{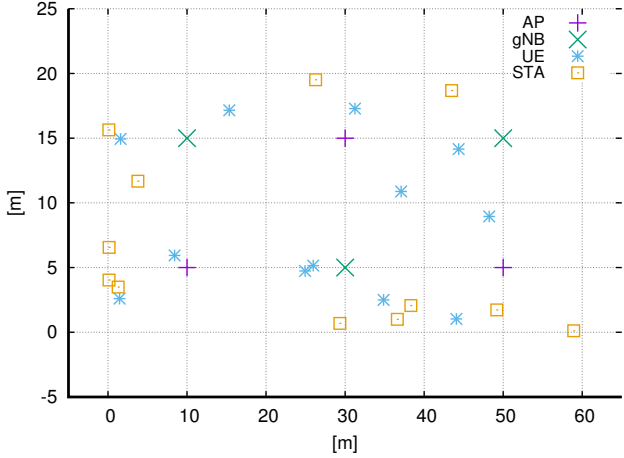}
  \caption{Indoor scenario with 3 gNBs, 3 APs, 12 UEs, and 12 STAs.}
  \label{fig:indoor}
\end{figure}

As a coexistence simulation scenario, we consider a dense indoor hotspot deployment shown in Figure~\ref{fig:indoor}, which is similar to the one evaluated by the 3GPP for coexistence between NR-U and Wi-Fi in the 5 GHz band. Since the coverage ranges at the mmWave frequencies would be shorter than the 5 GHz band, we used shorter maximum distance of 20 m between two devices, compared to 40 m in the 5 GHz band. The scenario shown in Figure~\ref{fig:indoor} consists of two operators deploying 3 {gNBs/APs} each, in a single floor building of 60 m $\times$ 20 m area. Each operator can deploy WiGig or NR-U technology and serves 12 users randomly distributed in the building. In this paper, we focus only on the standalone operation of NR-U, for being more challenging from the coexistence perspective. The remaining simulation parameters are given in Table~\ref{table:params}.

\subsection{Simulation campaign}\label{sec:uses_b}

We run simulation campaigns with different loads, but for space constraints we present here the results for a constant bit rate (50 Mbps) application for each device. This value is selected because it offers a high load, without reaching saturation, which allows us to observe interesting effects. In each plot, we depict the results for the following NR-U channel access mechanisms: 

\begin{itemize}
  \item \emph{On/On}: NR-U with AlwaysOn CAM at both the gNBs and the UEs;
  \item \emph{OnOff/OnOff}: NR-U with OnOff CAM with a 50$\%$ duty cycle of 9 ms, i.e., 9 ms ON and 9 ms OFF, at both the gNBs and the UEs;
  \item \emph{Cat4/On}: NR-U with Cat4 LBT at the gNBs and AlwaysOn at the UEs;
  \item \emph{Cat4/Cat2}: NR-U with Cat4 LBT at the gNBs and Cat2 LBT at the UEs;
  \item \emph{Cat3/On}: NR-U with Cat3 LBT at the gNBs and AlwaysOn at the UEs;
  \item \emph{Cat3/Cat2}: NR-U with Cat3 LBT at the gNBs and Cat2 LBT at the UEs. 
\end{itemize}

Approximately 120 seconds are needed to run a single simulation of 1.5 simulated seconds, but a parallel cluster can be used to increase the running number of simulations per hour. The output statistics presented in this paper are the channel occupancy, packet delay, and throughput. The results are meant to compare the base case, when both the operators deploy WiGig (denoted by \emph{WiGig only}), and when one operator deploys WiGig and the other deploys NR-U with different channel access mechanisms (described in Section~\ref{sec:uses_b}, with the channel access type indicated at the x-axis label). Each column represents an independent simulation set of 20 simulations with same parameters but different random seed. Visually, we have the maximum and minimum value plotted as whiskers, and the 95\% percentile and the 5\% percentile plotted as a box. In each box, a horizontal solid line represents the 50\% percentile.

\begin{table}[]
\footnotesize
\caption{Main scenario simulation parameters}
\centering
\label{table:params}
\begin{tabular}{l l}
\hline
Parameter          & Value       \\ 
\hline
\textbf{Deployment and configuration:}          &        \\ \hline
Channel model      & 3GPP Indoor Hotspot~\cite{TR38900}   \\ 
Channel bandwidth  & 2.16 GHz     \\ 
Central frequency  & 58 GHz      \\ 
Link adaptation    & Adaptive MCS    \\ 
gNB/AP antennas   & Uniform Planar Array 8x8     \\
UE/STA antennas     & Uniform Planar Array 4x4   \\
Transmission power  & 17 dBm \\
NR-U subcarrier spacing         & 120 kHz           \\ 
Noise power spectral density         & -174 dBm/Hz          \\
Noise figure         & 7 dB           \\ 
\hline
\textbf{NR-U LBT CAM:}          &        \\ \hline
gNB \gls{ed} threshold & -79 dBm (omniLBT)\\ 
UE \gls{ed} threshold & -69 dBm (dirLBT)\\ 
CCA slot duration & 5 us\\ 
defer interval during CCA  & 8 us\\
Maximum COT & 9 ms\\
Cat 4 LBT minimum CWS & 15 \\
Cat 4 LBT maximum CWS & 1023\\
Cat 3 LBT CWS & 15\\
Cat 2 LBT defer period & 25 us\\
\hline
\textbf{NR-U OnOff CAM:}          &        \\ \hline
duty cycle & 50\%: ON and OFF periods of 9 ms\\ 
\hline
\textbf{WiGig CSMA/CA:}          &        \\ \hline
AP/STA \gls{ed} threshold & -79 dBm (omniLBT)\\ 
CCA slot duration & 5 us\\ 
defer interval during CCA  & 8 us\\
CSMA/CA minimum CWS & 15 \\
CSMA/CA maximum CWS & 1023\\
\hline
\end{tabular}
\end{table}

\begin{figure*}
\centering
\subfloat[Channel Occupancy]{\includegraphics[width = 0.95\columnwidth]{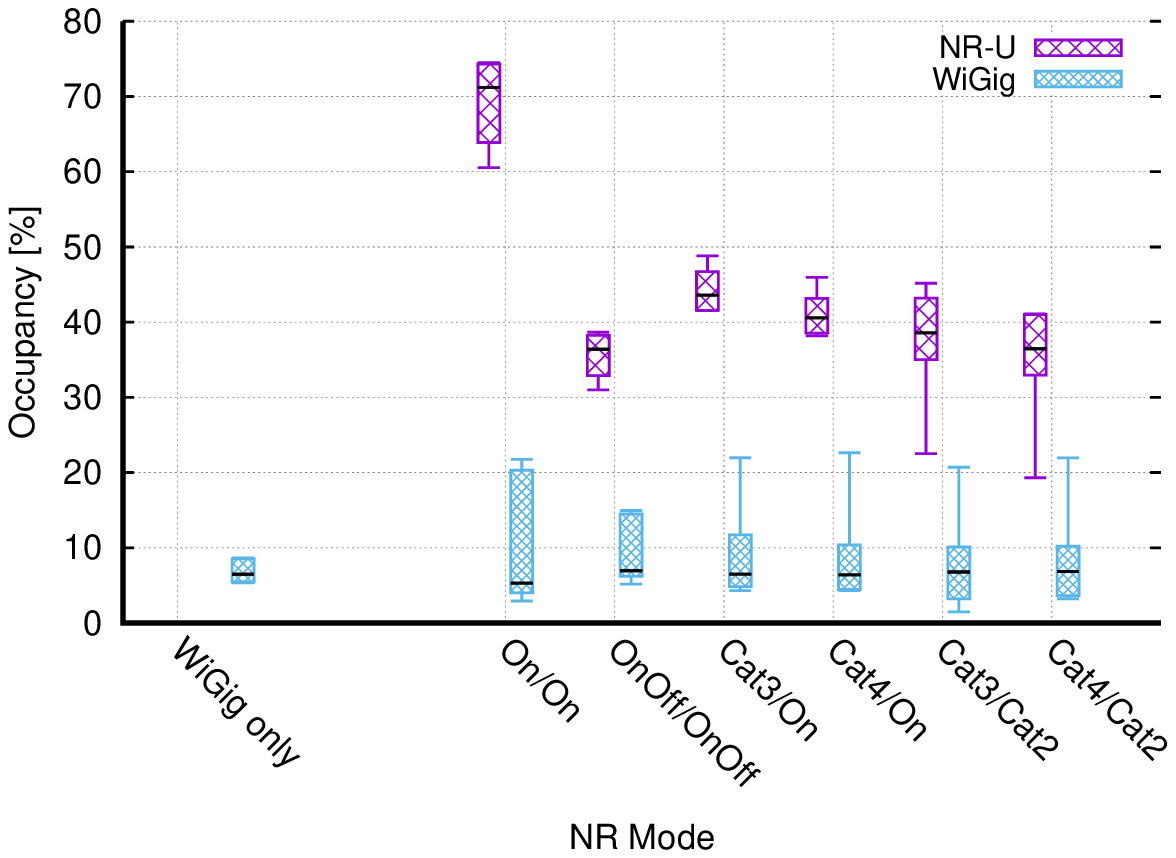}\label{fig:occupancy}}
\subfloat[E2E Latency]{\includegraphics[width = 0.95\columnwidth]{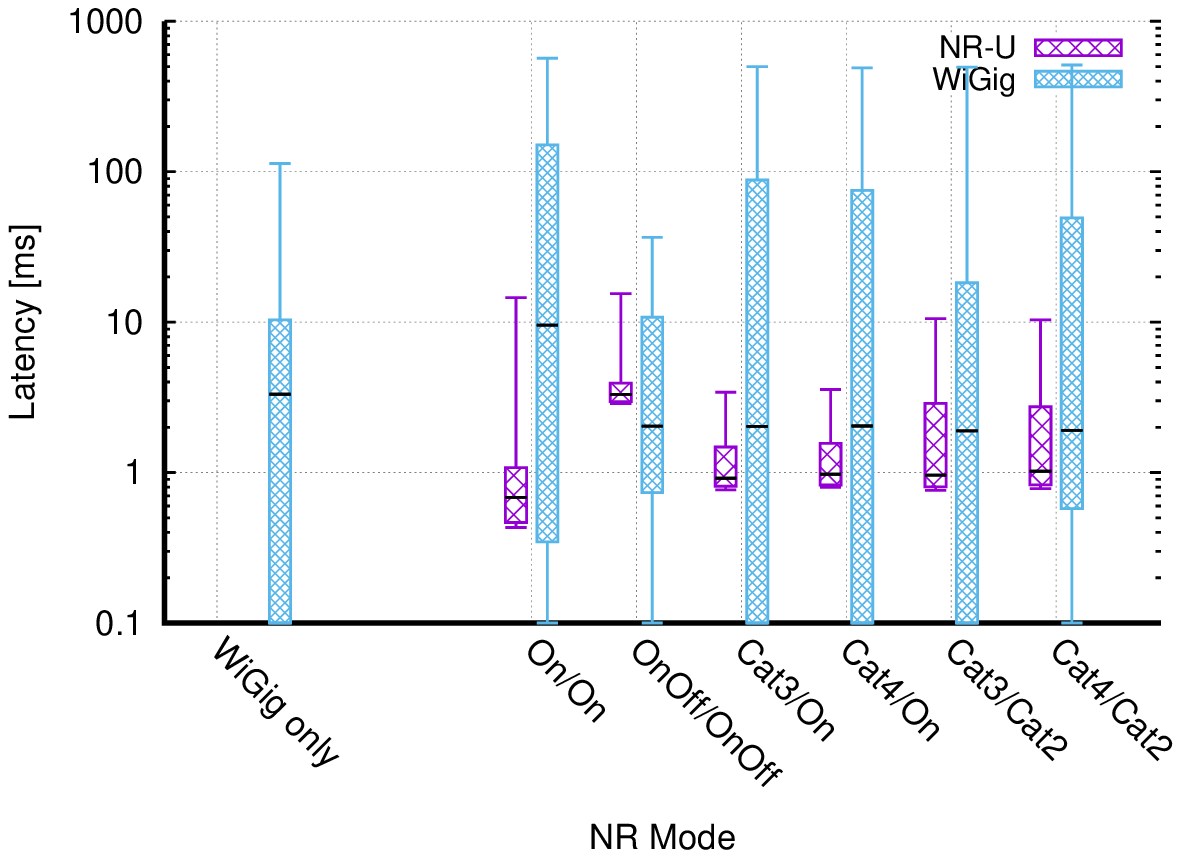}\label{fig:latency}}\\
\subfloat[E2E Goodput]{\includegraphics[width = 0.95\columnwidth]{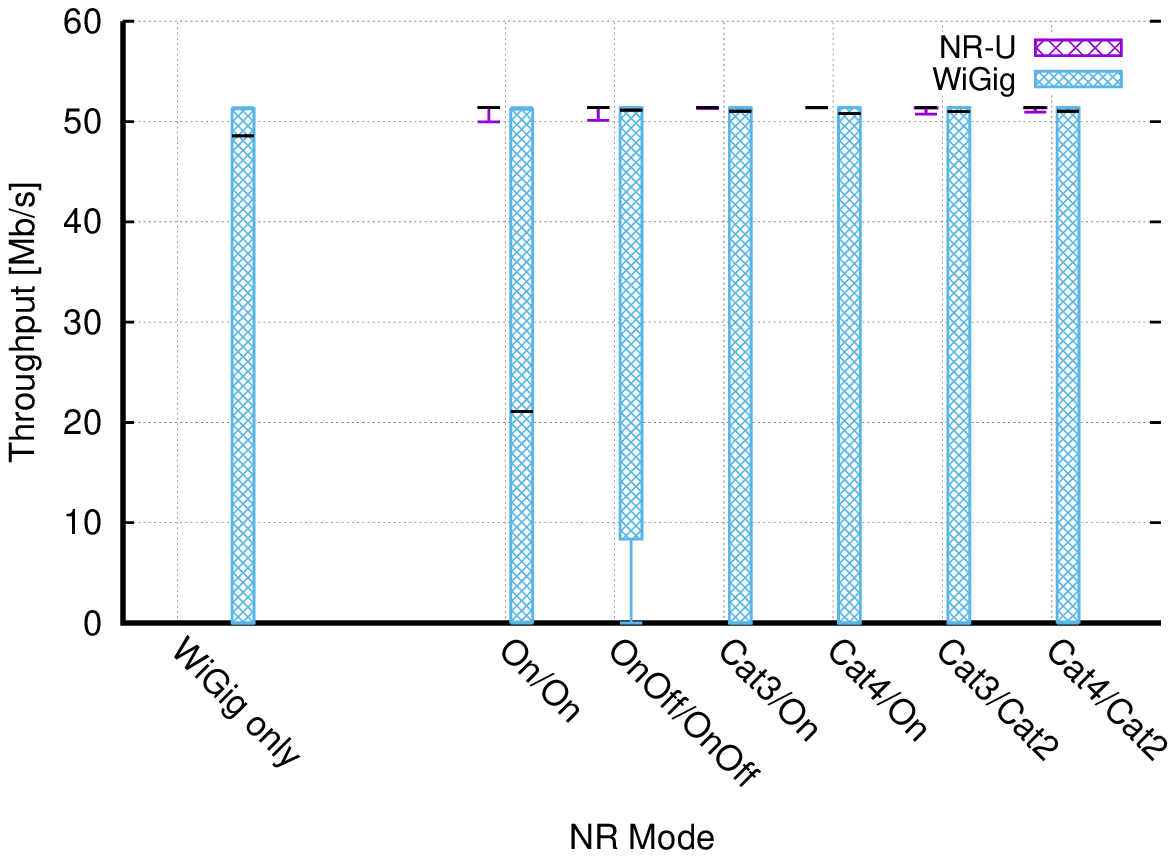}\label{fig:throughput}}\\
\caption{50 Mbps load per-device. The WiGig-only bar is obtained in which all the 24 users are IEEE 802.11ad-based, while for other bars, 12 devices are IEEE 802.11ad-based and the other half are using NR-U, with the channel access type indicated as the bar label.}
\label{fig:results}
\end{figure*}

\subsection{Simulation output}\label{sec:uses_c}

Based on the 3GPP fairness definition, NR-U is expected to operate in a \emph{fair} and \emph{friendly} manner to WiGig, by not impacting WiGig's performance more than another WiGig device would do~\cite{lagen:19}. In the following, we investigate each figure of merit highlighting interesting aspects and facts.

\textbf{Impact on channel occupancy.} As shown in Figure~\ref{fig:occupancy}, the channel occupancy of NR-U devices is significantly higher than the WiGig devices. The reason is that the minimum resource allocation granularity of NR-U is the entire OFDM symbol, while IEEE 802.11ad has no such restriction, and its channel occupancy strictly depends on the time needed to transmit the IEEE 802.11ad frames. Note that in WiGig, the duration of the transmissions varies for each IEEE 802.11ad frame, as a result of the selected \gls{mcs} and assuming that the whole bandwidth is used. From simulation data, the average length of a WiGig transmission is 3.5 us. In NR-U, when the traffic is low, the symbol remains partially empty, but the channel still occupies the whole symbol length (8.92 us for subcarrier spacing of 120 kHz), which leads to inefficient channel usage. This means that for the same data, an NR-U device is occupying the channel almost three times more than a WiGig device. A similar behavior was found in LTE-LAA and Wi-Fi coexistence~\cite{Bojovic2019}, where the difference was even higher, because of the higher LTE minimum allocation granularity.

On the other hand, {if} we compare the different NR-U channel access techniques, the channel occupancy of OnOff and LBT-based strategies (i.e., Cat1-Cat4) are lower than the AlwaysOn strategy. For the OnOff approach, the reason is that during the 9 ms OFF period, {a} NR-U {device} has time to accumulate data in \gls{rlc} buffers, {which enables} {to} fill the symbols {efficiently during} the transmission opportunities. Similarly, any LBT-based implementation {provides} more time to accumulate data during the sensing time, which{, however,} increases the delay (which is analyzed later in this section) but reduces the channel occupancy. 
Among the LBT {based techniques}, we observe that the {more} conservative the implementation {due to its ability to avoid collisions} is, the {more} the NR-U channel occupancy is reduced.

\textbf{Impact on latency.} Figure~\ref{fig:latency} shows that, {in terms of} delay, NR-U {performs} considerably better than IEEE 802.11ad. This is achieved thanks to NR-U specific features. On one hand, WiGig uses a contention-based access, which makes WiGig more prone to collisions. Instead, NR-U considers a slot-based access  and appropriate scheduling schemes, thus reducing collision probabilities. On the other hand, if transmissions collide or a blocking arises, \gls{harq} {in NR-U may} still successfully decode the frame through data recombination, while WiGig keeps retransmitting without combining until the maximum number of retransmissions is reached, thus eventually increasing the latency. We also observe that for these reasons, WiGig results have an higher standard deviation with respect to NR-U results.

In general, the WiGig's latency is reduced (compared to the WiGig-WiGig case) when the neighbour operator uses NR-U with LBT. The reason lies in the fact that a WiGig device uses a higher sensitivity threshold to listen the transmissions from non-WiGig devices compared to the transmission from other WiGig devices (due to preamble detection capabilities of WiGig nodes). Hence, a WiGig device backs off more often when coexisting with another WiGig than with another technology like NR-U.  
In turn, NR-U devices with LBT are better neighbors to WiGig devices than WiGig devices themselves. Comparing the different types of LBT, we can see that more conservative implementations (that consider LBT also at the UE side) increase the standard deviation of the end-to-end latency at NR-U devices, without substantial modifications in the WiGig latency performance. Unfortunately, without the LBT mechanism, the NR-U devices are not so friendly to WiGig, for the obvious reason that they are not listening to the channel before attempting a transmission, which often results in a collision.

\textbf{Impact on throughput.} Figure~\ref{fig:throughput} depicts the throughput results. We observe that in all cases the configurations can serve the offered traffic, so the saturation point of the system is not reached. The median of achieved throughout is always close to the highest value, except for the case in which WiGig coexists with NR-U AlwaysOn, where WiGig obtains a lower performance because it cannot find the channel free to transmit. In WiGig, minimum values of 0 Mbps are obtained because some STAs are getting interfered during the association phase and unable to associate with their AP. Moreover, we can see a higher variance for the WiGig nodes due to the same reasons highlighted before, for the latency results (i.e., quasi-omnidirectional based reception at AP in the uplink and the lack of retransmission combining). The throughput of some STAs is affected because the AP can miss the ACK feedback, interpreting it as a loss, generating unwanted retransmissions.

In all NR-U cases with different channel access mechanism, all the data can be delivered with extremely reduced standard deviation thanks to the directional reception and the use of HARQ combining. Moreover, the frame structure allows scheduling the UEs by the \gls{gnb}s, so that all the devices can be adequately served. Except for the case of AlwaysOn NR-U, NR-U does not have any adverse negative impact on throughput performance of the WiGig devices, which demonstrates that in terms throughput, NR-U with either duty cycle or any LBT based channel mechanism fulfills its coexistence design objective.

\textbf{Impact of CAM.} From the above results, we observe that all the channel access coexistence options for NR-U based on LBT or duty cycle (OnOff) are similarly friendly to WiGig. Thanks to the directionality of the transmissions and to the propagation conditions of the mmWave bands, the concrete LBT categories are not so determinant to the fairness, as it was for LTE-LAA, and the duty cycled implementation can also meet the fairness criterion. On the other hand, our preliminary results indicates that an AlwaysOn and uninterrupted implementation badly affects the WiGig nodes performance.

\section{Conclusions}
\label{sec:conc}
In this paper, we have presented the first open-source extension to the ns-3 simulator to perform coexistence studies between 3GPP and IEEE technologies in the unlicensed mmWave spectrum, from an end-to-end perspective. {For the 3GPP technology, an NR-U model based on an NR Release-15 implementation satisfying the 60 GHz regulatory requirements is considered.} We have investigated an indoor scenario with multiple users and a deployment of IEEE 802.11ad along with NR-U nodes in a single floor building. We have examined the impact of different NR-U channel access schemes (with LBT, without LBT, and with a duty-cycle pattern) on WiGig nodes, from different performance indicators. For channel occupancy, we observe that NR-U devices are occupying the channel longer than WiGig, however this is not affecting WiGig nodes in the proposed setup. Regarding the latency, considering that WiGig has a different sensitivity threshold towards NR-U nodes than to other WiGig nodes, we observe a sensible reduction in the WiGig back-off times when coexisting with NR-U as compared to WiGig, which results in favorable coexistence characteristics between the two technologies. Finally, we observe that both NR-U and WiGig can serve all the requested traffic when NR-U uses either LBT or duty-cycle mechanisms. We conclude that the directionality of transmissions and particular propagation patterns in the 60 GHz band, favor NR-U and WiGig coexistence, so that the fairness criterion is met, when NR-U uses either LBT or duty-cycle mechanisms. The absence of a specific coexistence oriented access instead, generates unfairness towards WiGig as it is to expect if NR-U operates in uninterrupted manner.

\section{Acknowledgments}

This work was partially funded by Spanish MINECO grant
TEC2017-88373-R (5G-REFINE) and Generalitat de Catalunya grant 2017 SGR 1195. Also, it was supported by InterDigital Communications, Inc. 

\bibliography{references}
\bibliographystyle{ieeetr}

\section*{Biographies}

\begin{IEEEbiographynophoto}{Natale Patriciello} received his MSc in Theoretical Computer Science from University of Bologna (Italy) in 2013, and his Ph.D. degree in ICT from UNIMORE, Italy, in 2017. He is researcher at CTTC, Barcelona, Spain.
\end{IEEEbiographynophoto}

\begin{IEEEbiographynophoto}{Sanjay Goyal} received 
his Ph.D. (2016) and M.S. (2012) from NYU Tandon School of Engineering. Currently, he is working with InterDigital Communications on next generation communication systems. 
\end{IEEEbiographynophoto}

\begin{IEEEbiographynophoto}{Sandra Lag\'en} received her M.S. and 
Ph.D. degrees from Universitat Polit\`ecnica de Catalunya (UPC), Spain, in 2013 and 2016, respectively. Since 2017, she is a Researcher at CTTC.  
\end{IEEEbiographynophoto}

\begin{IEEEbiographynophoto}{Lorenza Giupponi}
received her PhD from UPC, Spain, in 2007. She is a Senior Researcher in CTTC, and a member of the Executive Committee, where she acts as the Director of Institutional Relations.  
\end{IEEEbiographynophoto}

\begin{IEEEbiographynophoto}{Biljana Bojovi\'c} received her Electrical and Computer Engineering degree from the University of Novi Sad, Serbia, in 2008. She visited the Qualcomm Inst. at the University of California, San Diego. She is a researcher at CTTC.
\end{IEEEbiographynophoto}

\begin{IEEEbiographynophoto}{Alpaslan Demir} received his Ph.D. degree in Communications from Polytechnic Institute of NYU, US, in 2012. He is a Principal Engineer in the Future Wireless Business Unit at InterDigital Communications Corp. \end{IEEEbiographynophoto}
 
\begin{IEEEbiographynophoto}{Mihaela Beluri}
received her M.S. degree in electrical engineering from the Polytechnic University of Bucharest, Romania. She is currently a principal engineer with InterDigital Communications, working
on 5G technologies. 
\end{IEEEbiographynophoto}

\end{document}